\documentstyle[pre,aps,multicol,psfig]{revtex}
\newcommand{\ignore}[1]{} 

\begin{document}
\draft
\title{Spectral Correlations in the Crossover Transition from a
 Superposition of Harmonic Oscillators to the Gaussian
 Unitary Ensemble}

\author{Thomas Guhr$^1$ and Thomas Papenbrock$^{1,2}$}

\address{${}^1$ Max--Planck--Institut f\"ur Kernphysik, Postfach 103980,
         69029 Heidelberg, Germany}
\address{${}^2$ Institute for Nuclear Theory, Department of Physics, 
University of Washington, Seattle, WA 98195, USA}
\date{\today} 

\maketitle  

\begin{abstract}
  We compute the spectral correlation functions for the transition
  from a harmonic oscillator towards the Gaussian Unitary Ensemble
  (GUE). We use a variant of the supersymmetry method to obtain
  analytical results in a fast and elegant way.  In contrast to
  certain related transitions, the $k$--point correlation function
  possesses the $k\times k$ determinant structure of the GUE limit for
  the entire transition. The results are used to consider also the
  spectral correlations of a superposition of $M$ transition spectra.
  Our results are non--perturbative and are valid for all values of
  the transition parameter.
\end{abstract}


%\begin{multicols}{2}

%\narrowtext
\draft

\section{Introduction}
\label{sec1}

Random Matrix Theory~\cite{Mehta} is a powerful tool for the modeling
of spectral fluctuation properties.  Due to the general symmetry
constraints, a time-reversal invariant system with conserved or broken
rotation invariance is modeled by the Gaussian Orthogonal (GOE) or
Symplectic Ensemble (GSE), respectively, while the Gaussian Unitary
Ensemble (GUE) models the fluctuation properties of a system under
broken time-reversal invariance. These ensembles are known to describe
the generic fluctuation properties of chaotic quantum systems very
accurately~\cite{Bohigas,FH,MG,GMW}.  While numerical simulations of
the ensuing matrix models usually pose no serious difficulties, the
analytical calculations of the observables, i.e.~the correlation
functions, is generally a non-trivial task. In the case of the pure
ensembles, Mehta and Dyson~\cite{Mehta} solved the problem about
thirty years ago by introducing the orthogonal polynomial method.

However, since a generic physical system has, classically, a mixed
phase space, the spectral fluctuations of the corresponding quantum
system will be in between the pure cases. The transition from
preserved to broken time reversal invariance was worked out by Mehta
and Pandey~\cite{Mehta} in 1983. The transitions of the spectral
fluctuations in the case of gradually broken symmetries, i.e.~quantum
numbers was computed in Refs.~\cite{GW} and~\cite{Pandey95}.  Here we
will focus on a system that undergoes a transition from regular to
chaotic fluctuations.  To model such a system, we write the $N\times
N$ random matrix representing the total Hamiltonian as a sum of a
regular and a chaotic part,
\begin{equation}
H(\alpha) \ = \ H^{(0)} \, + \, \alpha H^{(1)}
\label{ha}
\end{equation} 
where $\alpha$ is the dimensionless transition parameter. The matrices
$H^{(1)}$ are drawn from a Gaussian Ensemble with the probability
density function $P_N^{(1)}(H^{(1)})$. Here, we are mainly interested
in the transition from a regular, equispaced spectrum to a chaotic
one.  This is a very important physical situation since many systems,
particularly in nuclear and molecular physics can be described as a
chaos producing part coupled to a harmonic oscillator. Often, it is
necessary to give the regular part $H^{(0)}$ a block structure which
reflects the presence of symmetries. This was already realized by
Porter and Rosenzweig~\cite{PoRo} who investigated, experimentally and
numerically, atomic spectra that contain various angular momentum and
spin quantum numbers subject to different coupling schemes.

However, at the moment, we make no assumptions yet for the probability
distribution $P_N^{(0)}(H^{(0)})$ of the matrices $H^{(0)}$.  The
decomposition~(\ref{ha}) can be justified for potential and billiard
systems.  Detailed numerical simulations for the transition of the
fluctuations can be found in Ref.~\cite{GWH}. However, despite several
attempts, full--fledged analytical discussions could only recently be
performed for the case of broken time reversal invariance in which the
matrices $H^{(1)}$ are drawn from the GUE. For a history of the
studies devoted to these problems see Ref.~\cite{GMW}. Presently,
there are the following techniques which make such calculations
possible.  Pandey~\cite{Pandey95} presented a certain construction of
the solution of Dyson's Brownian Motion Model~\cite{FH}. A related
approach was more recently put forward by Forrester~\cite{Forr}.  A
very direct and compact technique for the GUE was constructed in
Refs.~\cite{TGI,TGII}.  It relies on a variant of the supersymmetry
method~\cite{Efetov,VWZ} which was introduced in Ref.~\cite{TG}. The
enormous simplifications are due to the fact that supersymmetry can,
loosely speaking, be viewed as the ``irreducible representation'' of
Random Matrix Theory which becomes apparent in a new class of
diffusion equations~\cite{TGII,GMW}. Recently, Br\'ezin and
Hikami~\cite{BreHik} presented a third approach to derive similar
integral representations.

In this paper, we will apply the methods of Refs.~\cite{TGI,TGII} to
the transition starting from a harmonic oscillator.
Pandey~\cite{Pandey95} gave already a formula for the two-level
correlation function on the unfolded scale. Forrester~\cite{Forr}
extended the result for the $k$-level correlation function. Here, we
have three goals. First, we will show that this result can be obtained
very fast in a direct application of the general results of
Refs.~\cite{TGI,TGII}. Second, we will present plots and a detailed
discussion of the two-level correlation function. Third, we will go
beyond the known results and study a block structure of $H^{(0)}$ by
considering the superposition of $M$ transition spectra. After
briefly sketching the method in Sec.~\ref{sec2}, we work out the
crossover transition from one harmonic oscillator to the GUE in
Sec.~\ref{sec3}. In Sec.~\ref{sec4}, we study the superposition
of $M$ spectra. We summarize our results in Sec.~\ref{sec5}. 

\section{Transition ensembles}
\label{sec2}

Before turning to the harmonic oscillator, we briefly summarize the
general results. A detailed discussion can be found in
Ref.~\cite{TGII}, see also Ref.~\cite{GMG}.

As functions of the transition parameter $\alpha$, we wish to study
the $k$-level correlation functions 
\begin{eqnarray}
& &R_k(x_1,\ldots,x_k,\alpha) \ = \
      \frac{1}{\pi^k} \int d[H^{(0)}] \, P_N^{(0)}(H^{(0)}) \times
                                             \nonumber\\
& &\qquad \qquad  \int d[H^{(1)}] \, P_N^{(1)}(H^{(1)}) \,
      \prod_{p=1}^k {\rm Im}\,{\rm tr}\frac{1}{x_p^--H(\alpha)} 
\label{rkra}\end{eqnarray}
depending on $k$ energies $x_p, \ p=1,\ldots,k$.  where the energies
are given imaginary increments such that $x_p^\pm = x_p \pm
i\varepsilon$. It is convenient to work with correlation functions
$\widehat{R}_k(x_1,\ldots,x_k,\alpha)$ which involve the full Green
functions, including real and imaginary parts. By studying different
combinations of the signs of the imaginary parts of the Green
functions, we can construct the physically interesting
functions~(\ref{rkra}), see Ref.~\cite{TGII}.  The correlation
functions $\widehat{R}_k(x_1,\ldots,x_k,\alpha)$ that may be written
as the derivatives
\begin{equation}
\widehat{R}_k(x_1,\ldots,x_k,\alpha) \ = \ \frac{1}{(2\pi)^k} \,  
      \frac{\partial^k}{\prod_{p=1}^k \partial J_p} \,
      Z_k(x+J,\alpha) \Bigg|_{J=0}
\label{drka}\end{equation}
of the normalized generating functions $Z_k(x+J,\alpha)$. The energies
and the source variables are ordered in the diagonal matrices $x={\rm
  diag}(x_1,x_1,\ldots,x_k,x_k)$ and $J={\rm
  diag}(-J_1,+J_1,\ldots,-J_k,+J_k)$, respectively.  The desired
functions $R_k(x_1,\ldots,x_k,\alpha)$ can be derived from the
generating function $\Im Z_k$, where the symbol $\Im$ stands for the
proper linear combination~\cite{TGII}.  The average over the GUE is
done by means of the standard techniques of the supersymmetry
method~\cite{Efetov,VWZ}, yielding
\begin{eqnarray}
  & & Z_k(x+J,\alpha) \ = \ \int d[H^{(0)}] \, P_N^{(0)}(H^{(0)}) \int
  d[\sigma] \, \exp\left(-\frac{1}{\alpha^2}{\rm trg}\sigma^2\right)
  \times \nonumber\\
  & & \qquad \qquad {\rm
    detg}^{-1}\left((x^\pm+J-\sigma)\otimes 1_N - 1_{2k}\otimes
  H^{(0)}\right) \ .
\label{grkas}\end{eqnarray}
For details of the derivation and notation, the reader is referred to
Ref.~\cite{TG}.  In Eq.~(\ref{grkas}), $\sigma$ denotes a $2k\times
2k$ Hermitean supermatrix, and $1_N$ and $1_{2k}$ are $N \times N$ and
$2k \times 2k$ unit matrices, respectively.

To proceed, it is in our case advantageous to avoid the saddle point
approximation of Refs.~\cite{Efetov,VWZ}.  We shift the matrix $x+J$
is shifted from the graded determinant to the graded probability
density and the supermatrix $\sigma$ is diagonalized according to
$\sigma=u^{-1}su$, where $s={\rm
  diag}(s_{11},is_{12},\ldots,s_{k1},is_{k2})$. The volume element can
be rewritten as $d[\sigma]=B_k^2(s) d[s] d\mu(u)$ with
$B_k(s)=\det[1/(s_{p1}-is_{q2})]_{p,q=1,\ldots,k}$ the Jacobian, here
referred to as Berezinian.  The non--trivial integration over the
unitary diagonalizing supergroup with its Haar measure $d\mu(u)$ is
the crucial step and can be performed with the supersymmetric
extension~\cite{TG} of the Harish--Chandra Itzykson Zuber
integral~\cite{HCIZ}. Collecting everything we arrive at
\begin{eqnarray}
Z_k(x+J,\alpha) &=& 1-\eta(x+J) \, + \, \frac{1}{B_k(x+J)} 
                    \int G_k(s-x-J,\alpha) Z_k^{(0)}(s) B_k(s)
                    d[s] \ , \nonumber\\
Z_k^{(0)}(x+J) &=& \int d[H_0] P_N^{(0)}(H_0) {\rm detg}^{-1}
                 \left((x^\pm+J)\otimes 1_N - 1_{2k}\otimes H_0\right) \ ,
\label{zks}
\end{eqnarray}
where the kernel resulting from the group integration is Gaussian and
given by
\begin{equation}
G_k(s-r,\alpha) = \frac{1}{(\pi\alpha^2)^k}
          \exp\left(-\frac{1}{\alpha^2}{\rm trg}(s-r)^2\right) \ .
\label{izks}
\end{equation}
with $r=x+J$.  The distribution $1-\eta(x+J)$ in Eq.~(\ref{zks})
ensures the normalization $Z_k(x,\alpha)=1$ at $J=0$. This
distribution is not important for any of the formulae to follow, see
the discussion in Ref.~\onlinecite{TGII}. The generating function
$Z_k(x+J,\alpha)$ satisfies an exact diffusion equation in the curved
space of the eigenvalues of Hermitean supermatrices. Here,
$t=\alpha^2/2$ is the diffusion time and the generating function
$Z_k^{(0)}(x+J)$ serves as the initial condition.  This diffusion is
the supersymmetric analogue~\cite{TGII} of Dyson's Brownian
Motion~\cite{FH}.

The integration over $s$ requires to take a new type of boundary
contributions~\cite{Efetov,VWZ,Roth} into account which do not occur
in ordinary analysis. However, in Refs.~\cite{TG,TG4} it was shown
that we do not need to worry about them when calculating correlation
functions of the type we are interested in here.  Collecting
everything, we obtain the $k$--level correlation functions
\begin{eqnarray}
& & R_k(x_1,\ldots,x_k,\alpha) \ = \
                                             \nonumber\\ 
& & \qquad \qquad \frac{(-1)^k}{\pi^k} \, \int G_k(s-x,\alpha) \,
                  \Im Z_k^{(0)}(s) \, B_k(s)\,d[s]
\label{cfm}
\end{eqnarray}
for non-zero $\alpha$. The case $\alpha=0$ is trivial by construction.

As a last step it remains to unfold the correlation functions for
large level number $N$ by removing the dependence on the level
density.  We define new energies $\xi_p = x_p/D,\ p=1,\ldots,k$ in
units of the mean level spacing $D$. The transition parameter has to
be unfolded, too, $\lambda = \alpha/D$ and was first introduced by
Pandey~\cite{Pandey}.  The $k$-level correlation functions on the
unfolded scale $X_k(\xi_1,\ldots,\xi_k,\lambda) = \lim_{N\to\infty}
D^k R_k(x_1,\ldots,x_k,\alpha)$ are then generic, i.e.~translation
invariant over the spectrum.  It is useful to unfold the integration
variables $s$ in Eq.~(\ref{cfm}) by making the rescaling $s \to s/D$.
We arrive at
\begin{eqnarray}
& & X_k(\xi_1,\ldots,\xi_k,\lambda) \ = \ 
                                                  \nonumber\\
& & \qquad \qquad \frac{(-1)^k}{\pi^k} \, \int G_k(s-\xi,\lambda) \,
                  \Im z_k^{(0)}(s) \, B_k(s)d[s]
\label{cfum}
\end{eqnarray}
for non-zero $\lambda$ where the unfolded generating function of the
arbitrary correlations is given by
\begin{equation}
\label{smallz}
z_k^{(0)}(s) = \lim_{N\to\infty} Z_k^{(0)}(Ds) \ .  
\end{equation} 
Hence, we have expressed the unfolded $k$-level correlation function
for the transition from arbitrary to GUE fluctuations as a $2k$-fold
integral.

\section{Transition from an equispaced spectrum}
\label{sec3}

\begin{figure}[htb]
\begin{center}
\leavevmode
\parbox{\textwidth}{\psfig{file=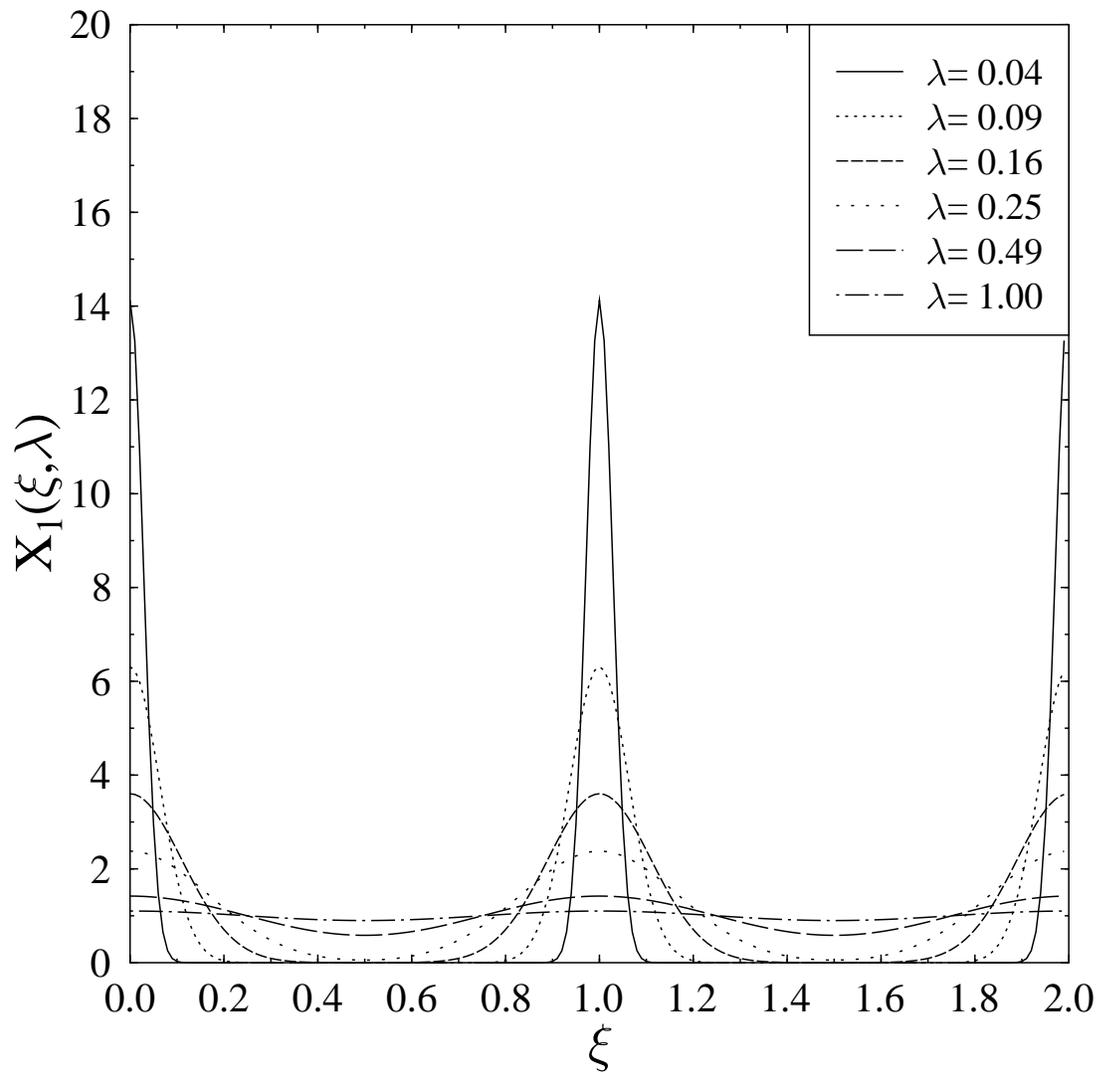,width=\textwidth,angle=0}}
\end{center}
\caption{Level density $X_1(\xi,\lambda)$ for different values of the 
transition parameter $\lambda$}
\label{dens}
\end{figure}

\begin{figure}[htb]
\begin{center}
\leavevmode
\parbox{\textwidth}{\psfig{file=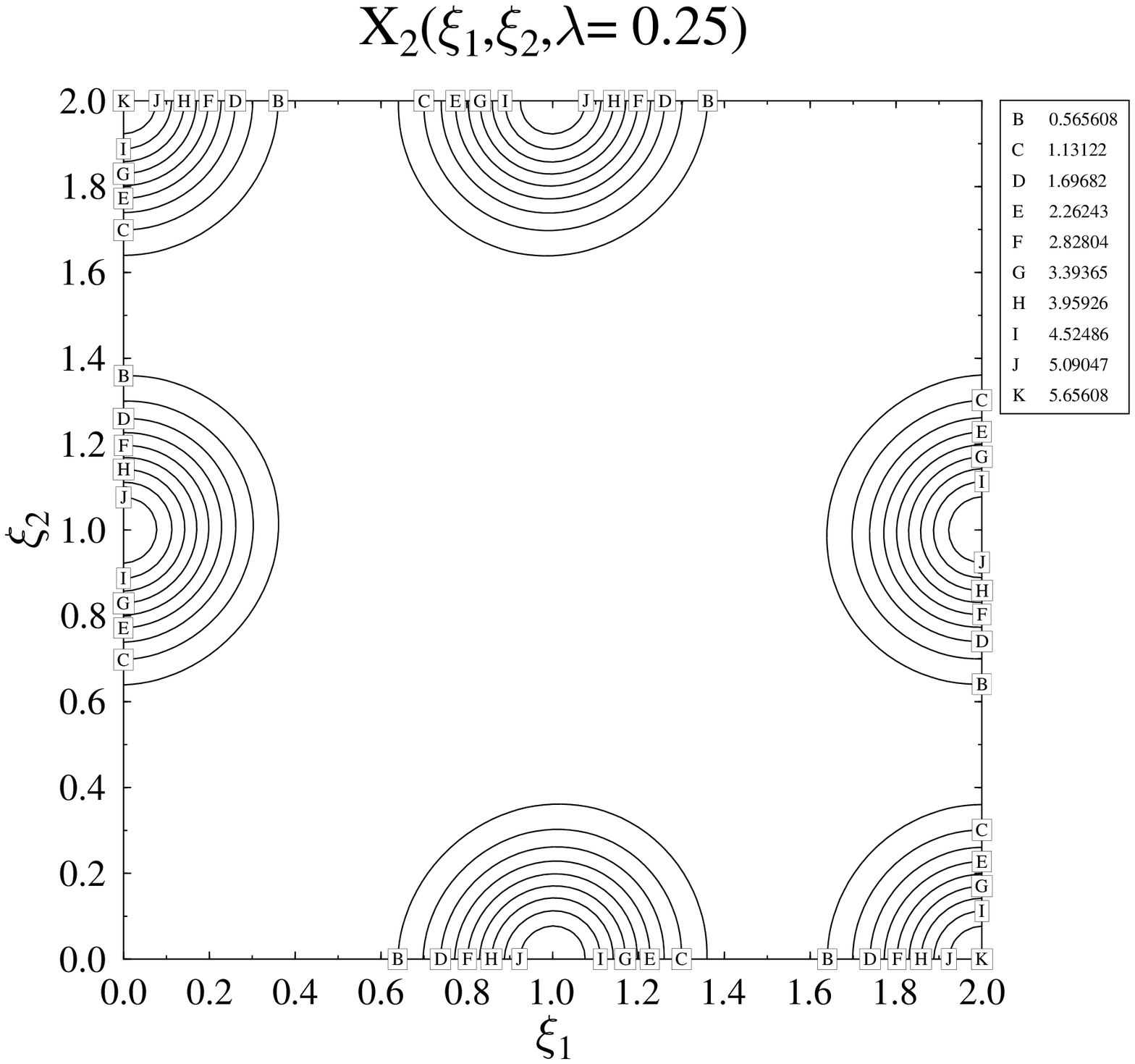,width=\textwidth,angle=0}}
\end{center}
\caption{Contour plot of $X_2(\xi_1,\xi_2,\lambda=0.25)$}
\label{2point1}
\end{figure}

\begin{figure}[htb]
\begin{center}
\leavevmode
\parbox{\textwidth}{\psfig{file=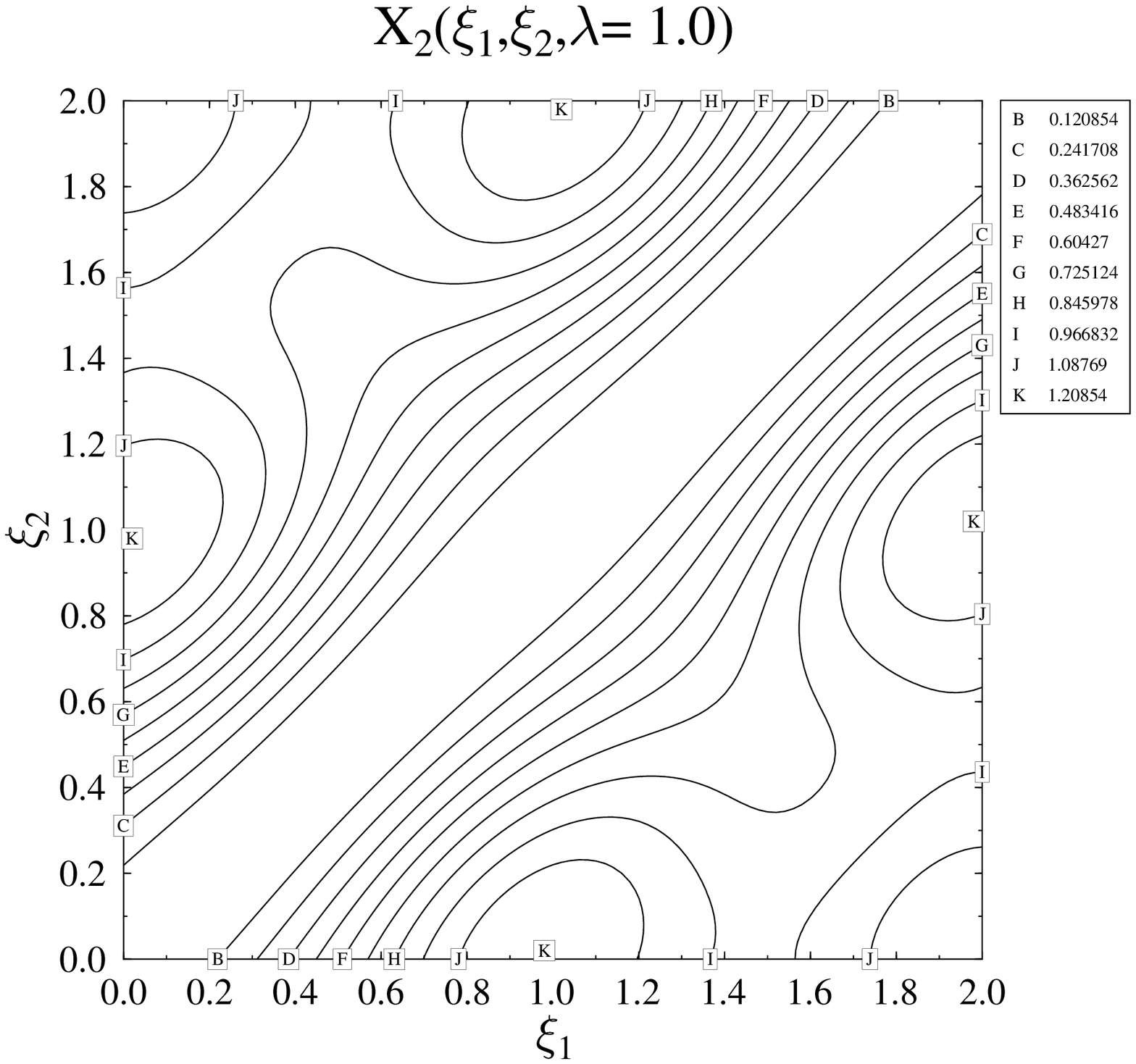,width=\textwidth,angle=0}}
\end{center}
\caption{Contour plot of $X_2(\xi_1,\xi_2,\lambda=1.0)$}
\label{2point2}
\end{figure}

All results derived so far are correct for arbitrary initial
correlations $R_k^{(0)}(x_1,\ldots,x_k)$ or
$X_k^{(0)}(\xi_1,\ldots,\xi_k)$. We now apply them to the case of a
equispaced spectrum. The harmonic oscillator probability density
function reads
\begin{eqnarray}
&& P_N^{(0)}(H^{(0)}) \ = \ \nonumber\\
&& \prod_{n=1}^N \delta(H_{nn}^{(0)}-\epsilon_n) 
   \prod_{n>m} \delta({\rm Re} H_{nm}^{(0)})
               \delta({\rm Im} H_{nm}^{(0)})
\label{ped}
\end{eqnarray}
where
\begin{equation}
\label{hospec}
\epsilon_n = \cases{(n-1)D/2 +\delta D, & $n=1,3,5,...,N$ \cr\cr
                    -nD/2 +\delta D,  & $n=2,4,6,...,N-1$ \ .}
\end{equation}
$D=2\sqrt{2/N}$ is the mean level spacing and the number of levels,
$N$, is odd. The spectrum is shifted by $\delta$ and we require
$|\delta| < 1$.  We find from Eq.~(\ref{zks}) for the generating
function of the initial condition
\begin{eqnarray}
  Z_k^{(0)}(s) =
  \prod_{p=1}^k\frac{is_{p_2}-\delta}{s_{p_1}^\pm-\delta}
  \prod_{n=1}^{(N-1)/2}
  \frac{\left(1-(is_{p_2}/(n+\delta)\right)
        \left(1+(is_{p_2}/(n-\delta)\right)}
       {\left(1-(is_{p_1}^\pm/(n+\delta)\right)
        \left(1+(is_{p_1}^\pm/(n-\delta)\right)}
\label{zkpuu}
\end{eqnarray}
According to Eq.~(\ref{smallz}) we evaluate the generating function on 
the unfolded scale in the limit $N\to\infty$ and obtain
\begin{equation}
z_k^{(0)}(s) \ = \ \prod_{p=1}^k\frac{\sin\pi( is_{p_2}-\delta)}
{\sin\pi( s_{p_1}^\pm-\delta)} \ .
\label{zkpu}
\end{equation}
The signs are determined by the choice of the sign of the imaginary
increment in the Green function.  The correlation functions can be
worked out by using the general result~(\ref{cfum}). The initial
condition takes the form
\begin{equation}
\label{gen}
\Im z_k^{(0)}(s) \ = \ \prod_{p=1}^k\sin{\pi( is_{p2}-\delta)}
      \, {\rm Im} \, \frac{1}{\sin\pi (s_{p1}^--\delta)} \ .
\end{equation}
We use the identity
\begin{equation}
{\rm Im} \ \frac{1}{\sin\pi( s^--\delta)} = 
\sum_{k=-\infty}^\infty(-1)^k\delta(s-k-\delta)
\end{equation}
and insert Eqs.~(\ref{izks}) and (\ref{gen}) into Eq.~(\ref{cfum}).
Since the function $B_k(s)$ is a determinant and since the generating
function of the harmonic oscillator spectrum (\ref{gen}) and the
Gaussian kernel (\ref{izks}) are products of $2k$ factors, the
$k$--point spectral correlation function may be written as a
determinant \cite{TG} 
\begin{equation}
\label{det}
X_k(\xi_1,\ldots,\xi_k,\lambda) = {\rm det}
\left[C(\xi_p,\xi_q,\lambda)\right]_{p,q=1\ldots k}
\end{equation} 
with 
\begin{eqnarray}
  C(\xi_p,\xi_q,\lambda) &=& -\frac{1}{\pi^2\lambda^2}
  \int\limits_{-\infty}^{+\infty}\int\limits_{-\infty}^{+\infty}
  \frac{ds_{p1}\,ds_{q2}}{s_{p1}-is_{q2}}\nonumber\\ 
  &&\times\exp{\left(\frac{1}{\lambda^2}
    \left((is_{q2}-\xi_q)^2-(s_{p1}-\xi_p)^2\right)\right)}
    \nonumber\\&&\times\sin{\pi( is_{q2}-\delta)}
    \sum_{l=-\infty}^{+\infty}(-1)^l\delta(s_{p1}-l-\delta)
    \nonumber\\ 
    &=&\frac{1}{\pi^2\lambda^2}\sum_{l=-\infty}^{+\infty}(-1)^l\,
    \exp\left(-\frac{(\xi_p-\delta-l)^2}{\lambda^2}\right)\nonumber\\ 
    &&\times\int\limits_{-\infty}^{+\infty} ds_{q2}\,\frac{\sin{\pi(
        is_{q2}-\delta)}}{s_{q2}+i(l+\delta)}\,
    \exp\left(-\frac{(s_{q2}+i\xi_q)^2}{\lambda^2}\right) \ .
\label{crep}
\end{eqnarray} 
The remaining integration may be done as follows. First, we translate
the path of integration about $+i\delta$ which does not change the
value of the integral. Then, we write the denominator of the integrand
as a Laplace transform and perform the integrations by means of
\cite{GR} 
\begin{eqnarray} 
  &&\int\limits_{-\infty}^{\infty}ds\frac{\sinh\pi s}{s+il}
  \exp\left(-\frac{\left(s+i(\xi-\delta)\right)^2}{\lambda^2}\right) =
  \nonumber\\ &&\quad-\pi (-1)^l
  \exp\left(\frac{(\xi-\delta-l)^2}{\lambda^2}\right) {\rm Im}\, {\rm
    erf}\left(\frac{l-(\xi-\delta)}{\lambda} \text{sign}(l)
  -i\frac{\pi}{2}\lambda\right) \ .
\label{zb1}
\end{eqnarray} 
This yields (${\rm erf}$ denotes the error function as defined in
Ref.~\cite{GR})
\begin{eqnarray}
\label{q}
C(\xi_p,\xi_q,\lambda) &=& -\frac{1}{\pi\lambda^2}
\exp{\left(-\frac{(\xi_p-\delta)^2-(\xi_q-\delta)^2}{\lambda^2}\right)}\nonumber\\ 
&&\times \sum_{l=-\infty}^\infty
\exp\left(-\frac{2}{\lambda^2}(\xi_q-\xi_p)l\right) \nonumber\\ 
&&\times{\rm Im} \, {\rm erf}\left(\frac{l-(\xi_q-\delta)}{\lambda} \ 
\text{sign}(l) -i\frac{\pi}{2}\lambda\right) \ .
\end{eqnarray}
To proceed, we use the definition of the error function and write it
as the integral of a Gaussian. Taking its imaginary part and writing
the just introduced Gaussian as a Fourier transform yields
\begin{eqnarray}
  &&C(\xi_p,\xi_q,\lambda) = \nonumber\\ &&\qquad\frac{1}{4\pi}
  \int\limits_{-1}^1dt \ \exp\left(\frac{\pi^2\lambda^2 t^2}{4}\right)
  \int\limits_{-\infty}^{\infty}ds \ \exp\left(-\frac{\lambda^2
    s^2}{4}\right) \nonumber\\ &&\qquad\times\sum_{l=-\infty}^{\infty}
  \cos\left(\pi(\xi_q-\delta-l)t\right)\cos\left((\xi_p-\delta-l)s\right)
  \ .
\label{zb2}
\end{eqnarray}
We use Poisson's sum formula and perform the remaining integrations.
The final result displays $C(\xi_p,\xi_q,\lambda)$ as a Fourier series
\begin{eqnarray}
  &&C(\xi_p,\xi_q,\lambda)=\sum_{l=-\infty}^\infty
  \exp{\left(-\pi^2\lambda^2l^2\right)}\, \nonumber\\ 
  &&\quad\times{\rm Re} \left(\exp\left(i2\pi l(\xi_p-\delta)\right)
  \frac{\sinh\left(\pi^2\lambda^2l+i\pi(\xi_q-\xi_p)\right)}
  {\pi^2\lambda^2l+i\pi(\xi_q-\xi_p)}\right) \ ,
\label{zb3}
\end{eqnarray}
which inserted in Eq.~(\ref{det}) completes our calculation.
Two interesting limits may be considered.
In the limit of vanishing transition parameter one obtains
\begin{equation}
\lim_{\lambda\to 0} C(\xi_p,\xi_q,\lambda) = 
\frac{\sin\pi\left(\xi_q-\xi_p\right)}{\pi\left(\xi_q-\xi_p\right)}
\sum_{l=-\infty}^\infty\delta(\xi_p-\delta-l) \ ,
\end{equation}
which generates the spectral correlations of the harmonic oscillator.
The opposite limit of infinite transition parameter yields the GUE
spectral correlations, i.e.
\begin{equation}
\label{GUE}
\lim_{\lambda\to\infty} C(\xi_p,\xi_q,\lambda) =
\frac{\sin\pi\left(\xi_p-\xi_q\right)} {\pi\left(\xi_p-\xi_q\right)}
\ .
\end{equation} 
Plots of the level density 
\begin{equation}
\label{onepoint}
X_1(\xi,\lambda)=C(\xi,\xi,\lambda)=\sum_{l=-\infty}^\infty 
\exp{\left(-\pi^2\lambda^2l^2\right)}\,
\frac{\sinh\pi^2\lambda^2l}{\pi^2\lambda^2l}\cos 2\pi l(\xi-\delta)
\end{equation}
are shown for different values of the transition parameter $\lambda$
in Fig.~\ref{dens}. Note that very small values of the transition
parameter $\lambda$ already cause a considerable broadening of the
$\delta$--peaks of the original harmonic oscillator spectrum.  It has
been observed in almost all crossover transitions that, on scales of a
few mean level spacings, the GUE limit is reached when $\lambda$ is of
the order of unity, see the review in Ref.~\cite{GMW}.
Figs.~\ref{2point1} and~\ref{2point2} show that this is also true for
the two--point function
\begin{equation}
\label{2point}
X_2(\xi_1,\xi_2,\lambda)=X_1(\xi_1,\lambda)X_1(\xi_2,\lambda)
-C(\xi_1,\xi_2,\lambda) \, C(\xi_2,\xi_1,\lambda) \ . 
\end{equation}
The delta--peaks at integer values $\xi_1\neq\xi_2$ are broadened for
a small value of the transition parameter $\lambda$, and the
GUE--correlations are approached fast.  Note also that the two--point
function (\ref{2point}) becomes translation invariant (i.e. it depends
only on the distance $\xi_1-\xi_2$) if averaging over $\delta$ is done
\cite{Forr}.

\section{Superposition of spectra}
\label{sec4}

We now consider a superposition of independent, non--interacting
spectra as the initial condition. This is equivalent to saying that
there is a symmetry that allows one to write the initial condition in
the form $H^{(0)} = {\rm diag}(H_1^{(0)},\ldots,H_M^{(0)})$ where each
of the $N_m \times N_m$ matrices $H_m^{(0)}, m=1,\ldots,M$ is drawn
from an independent ensemble. We have $\sum_{m=1}^M N_m = N$, the
total dimension, and $\sum_{m=1}^M D_m^{-1} = D^{-1}$, where $D_m$ is
the mean level spacing in the $m^{\rm th}$ ensemble. The quantities
$g_m=D/D_m$ are referred to as fractional level densities.  If all
$N_m$ are equal, we have $g_m=1/M$.  Symmetries which lead to such
models are frequently found in nuclear, atomic and molecular
physics~\cite{PoRo,GMW}.

For given fluctuations of every sub--ensemble with index $m$, the
fluctuations properties of the superposition can be worked out in a
rather straightforward way.  In the case of the nearest neighbor
spacing distribution, the result is given in the article by Porter and
Rosenzweig \cite{PoRo}.  More general discussion may be found in
Ref.~\cite{Mehta} and in the review~\cite{GMW}.

Here, we address the case that all matrices $H_m^{(0)}$ are drawn from
harmonic oscillators.  In discussing the influence of a chaotic
admixture to this initial condition, two different scenarios are of
interest: The chaotic admixture can either preserve the symmetry or
break it.  In the first scenario, we have to add a GUE matrix to {\it
  every} matrix $H_m^{(0)}$. Thus, we still have a superposition of
independent spectra and the total matrix $H(\alpha)$ still has the
block structure, i.e.~$H(\alpha) = {\rm
  diag}(H_1(\alpha),\ldots,H_M(\alpha))$.

In the second scenario, we add one $N\times N$ GUE matrix to the block
diagonal matrix $H^{(0)}$. Thus, for non--zero transition parameter,
the total matrix $H(\alpha)$ has no block structure. This scenario is
in the spirit of the Porter--Rosenzweig model~\cite{PoRo}. For $M=2$
such a symmetry breaking was investigated in Refs.~\cite{GW}
and~\cite{Pandey95}. In these studies, however, and in contrast to the
present one, the initial condition was also chaotic.  We discuss the
first scenario in Sec.~\ref{sec41}, before we briefly turn to the
second one in Sec.~\ref{sec42}.

\subsection{Symmetry preserving case}
\label{sec41}

Each of the block matrices is a harmonic oscillator, coupled to a GUE.
For simplicity, we assume that all blocks matrices shall have the same
dimension $N_m$ and the same coupling to the GUE. Moreover, to lift
the degeneracies, the $m^{\rm th}$ oscillator spectrum is chosen with
an energy shift $\delta_m=m/M$ (cf.  Eq.(\ref{hospec})). We are
interested in the unfolded two--point correlation function
\begin{eqnarray}
\label{super2}
X_2^{(M)}(\xi_1,\xi_2,\lambda)&=&
\sum_{m=1}^{M}g_m^2X_{2,m}(g_m\xi_1,g_m\xi_2,g_m\lambda)\nonumber\\ 
&+&\sum_{n,m=1}^Mg_mg_nX_{1,m}(g_m\xi_1,g_m\lambda)
X_{1,n}(g_n\xi_2,g_n\lambda)\nonumber\\ 
&-&\sum_{m=1}^Mg_m^2X_{1,m}(g_m\xi_1,g_m\lambda)
X_{1,m}(g_m\xi_2,g_m\lambda) \ .  
\end{eqnarray} 
where $g_m=1/M$ is the fractional density of each block matrix.  To
motivate this equation, we notice that
$X_2^{(M)}(\xi_1,\xi_2,\lambda)$ is the probability density of finding
a level at $\xi_1$ and another one at $\xi_2$. The levels can either
belong to the same symmetry sector (first line of Eq. (\ref{super2}))
or to different symmetry sectors (second and third line of Eq.
(\ref{super2})). Since the second line includes also correlations
within one symmetry sector (these have been accounted for in the first
line), the third line must be subtracted.  We emphasize once more that
the functions $X_{1,m}(g_m\xi_1,g_m\lambda)$ are not unity in the case
of a harmonic oscillator.

\begin{figure}[htb]
\begin{center}
\leavevmode
\parbox{\textwidth}{\psfig{file=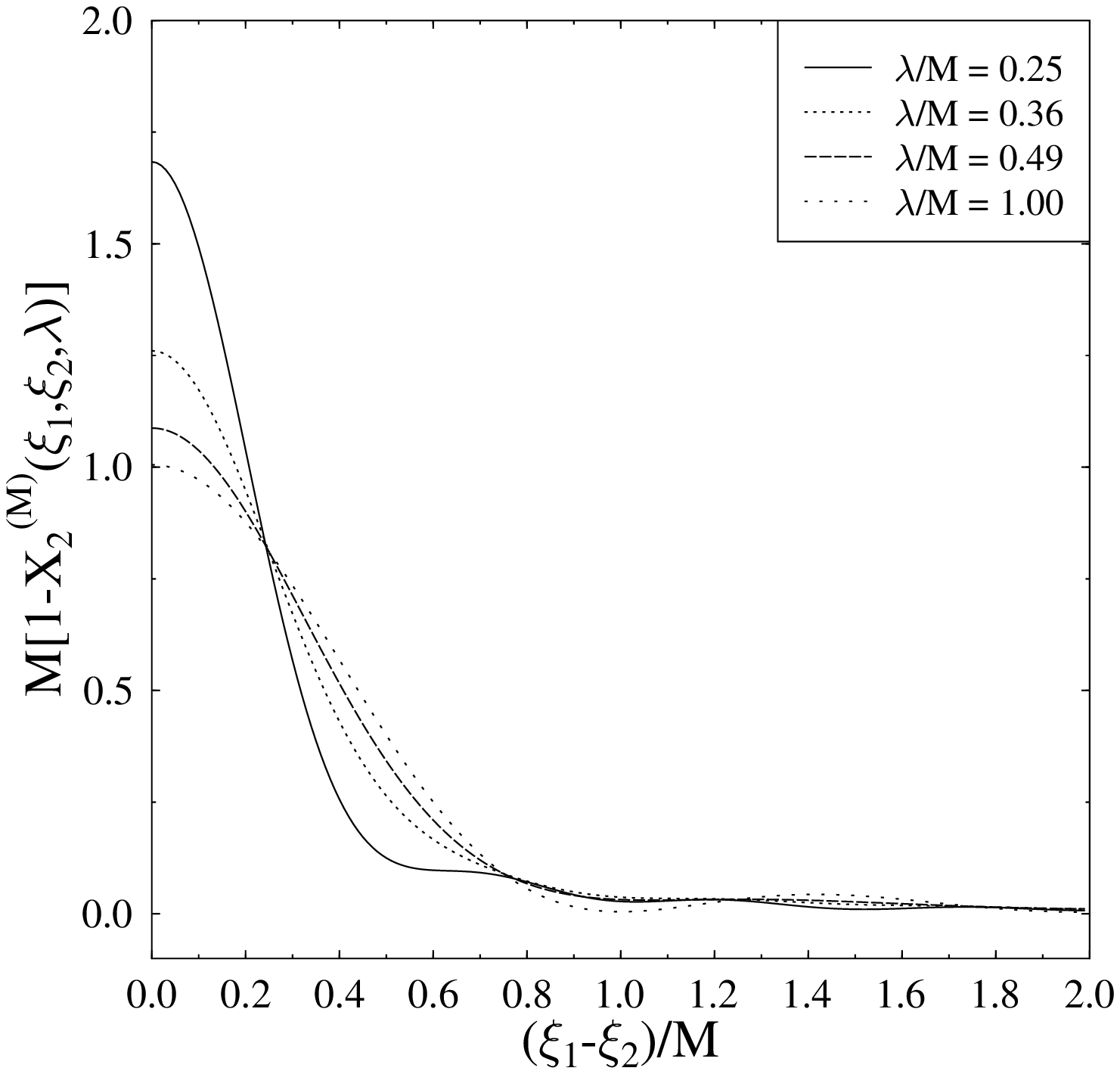,width=\textwidth,angle=0}}
\end{center}
\caption{Plot of $M[1-X_2^{(M)}(\xi_1,\xi_2,\lambda)]$ for different 
values of the transition parameter $\lambda$.}
\label{superplot}
\end{figure}

We now insert the the two--point function (\ref{2point}) with the
appropriate energy shifts into Eq.~(\ref{super2}) and interchange the
summation over $m$ with the summation over $k,l$. For $M >1$, this
yields expressions like
\begin{equation}
\sum_{m=1}^M\cos
2\pi\frac{k}{M}(\xi_1-m)\,\cos 2\pi\frac{l}{M}(\xi_2-m) =
\frac{M}{2}\left(\delta_{k,l}+\delta_{k,-l}\right)\cos
2\pi\frac{k}{M}(\xi_1-\xi_2) 
\label{zb4}
\end{equation} 
which help to perform one more summation. The final result reads for
$M > 1$
\begin{eqnarray}
\label{super2p}
X_2^{(M)}(\xi_1,\xi_2,\lambda)=
1-\frac{1}{M}\sum_{k=-\infty}^{\infty}&&\!\!\!\!\!\!\!
\left|\frac{\sinh\left(\pi^2\lambda^2k/M^2+i\pi(\xi_1-\xi_2)/M\right)}
  {\pi^2\lambda^2 k/M^2+i\pi(\xi_1-\xi_2)/M}\right|^2 \nonumber\\ 
  &\times&\exp\left(-2\pi^2\lambda^2\frac{k^2}{M^2}\right)\cos
  2\pi\frac{k}{M}(\xi_1-\xi_2) \ .
\end{eqnarray} 
In the limit of vanishing $\lambda$ we find
\begin{equation}
\lim_{\lambda\to 0}X_2^{(M)}(\xi_1,\xi_2,\lambda)=
   1-\delta(\xi_1-\xi_2) \ ,
\label{zb5}
\end{equation} 
and in the limit of infinite transition parameter we recover the
two--point correlation function for a superposition of $M$ GUE's 
\begin{equation}
\label{zb6}
\lim_{\lambda\to\infty}X_2^{(M)}(\xi_1,\xi_2,\lambda)=1
  -\frac{1}{M}\left(\frac{\sin\pi(\xi_1-\xi_2)/M}
                         {\pi(\xi_1-\xi_2)/M}\right)^2 \ .
\end{equation} 
For large $M$ and nonzero transition parameter $\lambda$, the
two--point correlation function approaches the value one.  Thus, the
spectrum becomes completely uncorrelated and exhibits the same
correlations as the (randomly generated) Poisson--spectrum. We also
note, that the two--point correlation function (\ref{super2p}) depends
only on the difference $(\xi_1-\xi_2)$ of its arguments.  This
translation invariance results from the assumption that the $m^{\rm
  th}$ spectrum is shifted by $m/M$ mean level spacings. Thus, any
deviation from translation invariance gives a hint to clustering or
degeneracy of levels belonging to different symmetry sectors.
Figure~\ref{superplot} shows the a plot of
$M[1-X_2^{(M)}(\xi_1,\xi_2,\lambda)]$ for various values of the
coupling $\lambda$. This function depends only on the ratios
$\lambda/M$ and $(\xi_1-\xi_2)/M$.

\subsection{Symmetry breaking case}
\label{sec42}

We turn to the second scenario defined above. We do this briefly,
because we mainly want to show the very different structure of
the result. In general, the initial condition on the original
scale takes the form
\begin{equation}
Z_k^{(0)}(s) = \prod_{m=1}^M Z_{km}^{(0)}(s)
\label{zb7}
\end{equation}
where the initial conditions $Z_{km}^{(0)}(s)$ are still completely
arbitrary. On the unfolded scale, we have
\begin{equation}
z_k^{(0)}(s) = \prod_{m=1}^M z_{km}^{(0)}(g_ms) \ .
\label{zb8}
\end{equation}
These formulae have to be inserted into Eqs.~(\ref{cfm}) 
and~(\ref{cfum}), respectively, to obtain the correlation functions.

If all $M$ initial ensembles are chosen as harmonic oscillators, we
find from Eq.~(\ref{zkpu}) on the unfolded scale
\begin{equation}
  z_k^{(0)}(s) \ = \ \prod_{m=1}^M \prod_{p=1}^k\frac{\sin\pi (g_m
    is_{p_2}-\delta_m)} {\sin\pi( g_m s_{p_1}^\pm-\delta_m)} \ .
\label{zb9}
\end{equation}
Since this is still a product, the determinant structure~(\ref{det})
of the $k$--point correlation functions is still preserved. The
functions 
\begin{eqnarray}
  C(\xi_p,\xi_q,\lambda) &=& -\frac{1}{\pi^2\lambda^2}
  \int\limits_{-\infty}^{+\infty}\int\limits_{-\infty}^{+\infty}
  \frac{ds_{p1}\,ds_{q2}}{s_{p1}-is_{q2}}\nonumber\\ 
  &&\times\exp{\left(\frac{1}{\lambda^2}
    \left((is_{q2}-\xi_q)^2-(s_{p1}-\xi_p)^2\right)\right)}
    \nonumber\\ 
  &&\times\prod_{m=1}^M\sin{\pi( g_m is_{q2}-\delta_m)}
    \ {\rm Im} \prod_{m^\prime =1}^M\frac{1}{\sin{\pi( g_{m^\prime}
        s_{q1}^--\delta_{m^\prime})}} \ .
\label{creps}
\end{eqnarray} 
are the entries of this determinant. For general $g_m$ and $\delta_m$,
this result is not easily amenable to simplifications. In some special
cases, however, the product over $m$ in Eq.~(\ref{zb9}) can be
performed.  In particular, for $g_m=1/M$ and $\delta_m=m/M$ we obtain
\begin{equation}
  z_k^{(0)}(s) \ = \ \prod_{p=1}^k
                     \frac{\sin\pi is_{p_2}} 
                          {\sin\pi s_{p_1}^\pm}
\label{zb10}
\end{equation}
by using standard results for trigonometric functions. This is
precisely Eq.~(\ref{zkpu}) for $\delta=0$. 
Thus, we recover the results of Sec.~\ref{sec3} because the
case discussed here corresponds to the coupling of one oscillator,
with the spacing on the original scale multiplied by $1/M$, to one
GUE--matrix.

\section{Summary}
\label{sec5}

We have computed all spectral correlation functions for the transition
from a harmonic oscillator to the GUE. We have used these results to
compute the two--point correlation function for a system which
undergoes a crossover transition to a GUE starting from a
superposition of independent oscillator spectra. We discussed chaotic
admixtures to an initial block structure which either preserve or
break this symmetry.  These ideas and techniques can also be used for
other block diagonal initial conditions with different individual
statistics.

Applying a variant of the supersymmetry technique, we gave an elegant
and very straightforward derivation of these results.

\section*{Acknowledgment}

TG acknowledges support of the Heisenberg Foundation.

%\end{multicols}

\begin{references}
\bibitem{Mehta}       M.L. Mehta, {\it Random Matrices}, 2nd ed.
                      (Academic Press, New York-London, 1991).
\bibitem{Bohigas}     O. Bohigas, 
                      in {\em Chaos and Quantum Physics},
                      edited by M.-J. Giannoni {\em et al.} 
                      (North--Holland, Amsterdam, 1991)
\bibitem{FH}          F. Haake, {\it Quantum Signatures of Chaos}
                      (Springer, Berlin, 1991).
\bibitem{MG}          M.C. Gutzwiller, {\it Chaos in Classical and
                      Quantum Mechanics} (Springer, New York, 1990).
\bibitem{GMW}         T. Guhr, A. M\"uller--Groeling and H.A. Weidenm\"uller,
                      Phys. Rep. {\bf 299} 189, (1998). 
\bibitem{GW}          T. Guhr and H.A. Weidenm\"uller, 
                      Ann. Phys. (NY) {\bf 199}, 412 (1990).
\bibitem{Pandey95}    A. Pandey, Chaos, Solitons and Fractals 
                      {\bf 5}, 1275 (1995).
\bibitem{PoRo}        Porter and Rosenzweig, 
                      Phys. Rev. {\bf 120} (1960) 1698 
\bibitem{GWH}         T. Guhr and H.A. Weidenm\"uller, Ann. Phys. (NY)
                      {\bf 193}, 472 (1989).
\bibitem{Forr}        P.J. Forrester, Physica {\bf A 223}, 365 (1996)
\bibitem{TGI}         T. Guhr,
                      Phys. Rev. Lett. {\bf 76}, 2258 (1996).
\bibitem{TGII}        T. Guhr,
                      Ann. Phys. (NY) {\bf 250}, 145 (1996)
\bibitem{Efetov}      K.B. Efetov, Adv. Phys. {\bf 32}, 53 (1983).
\bibitem{VWZ}         J.J.M. Verbaarschot, H.A. Weidenm\"uller and
                      M.R. Zirnbauer, 
                      Phys. Rep. {\bf 129}, 367 (1985).
\bibitem{TG}          T. Guhr, 
                      J. Math. Phys. {\bf 32}, 336 (1991).
\bibitem{BreHik}      E. Br\'ezin and S. Hikami,
                      Nucl. Phys. {\bf B479}, 697 (1996);
                      Phys. Rev. {\bf E56} 264 (1997).
\bibitem{GMG}         T. Guhr and A. M\"uller--Groeling,
                      J. Math. Phys. {\bf 38}, 1870 (1997).
\bibitem{HCIZ}        Harish-Chandra, Amer. J. Math. 
                      {\bf 80}, 241 (1958);
                      C. Itzykson and J.B. Zuber, 
                      J. Math. Phys. {\bf 21}, 411 (1980).
\bibitem{Roth}        M.J. Rothstein, 
                      Transactions of the American Mathematical Society 
                      {\bf 299}, 387 (1987).
\bibitem{TG4}         T. Guhr, 
                      Nuc. Phys. {\bf A560}, 223 (1993).
\bibitem{Pandey}      A. Pandey, 
                      Ann. Phys. (NY) {\bf 134}, 110 (1981).
\bibitem{GR}          I.S. Gradshteyn and I.M. Ryzhik,
                      {\it Table of Integrals, Series and Products}
                      (Academic Press, San Diego, 1980),
                      formula 3.897
\end{references}
\end{document}